# Bending-wave Instability of a Vortex Ring in a Trapped Bose-Einstein Condensate

T.-L. Horng,[1] S.-C. Gou,[2] and T.-C. Lin[3]

[1]*Department of Applied Mathematics, Feng Chia University, Taichung 40074, Taiwan*
[2]*Department of Physics, National Changhua University of Education, Changhua 50058, Taiwan*
[3]*Department of Mathematics, National Taiwan University, Taipei 10617, Taiwan*
(Dated: April 1, 2006)

Based on a velocity formula derived by matched asymptotic expansion, we investigate the dynamics of a circular vortex ring in an axisymmetric Bose-Einstein condensate in the Thomas-Fermi limit. The trajectory for an axisymmetrically placed and oriented vortex ring is entirely determined, revealing that the vortex ring generally precesses in condensate. The linear instability due to bending waves is investigated both numerically and analytically. General stability boundaries for various perturbed wavenumbers are computed. In particular, the excitation spectrum and the absolutely stable region for the static ring are analytically determined.

Vortices are fundamental excitations occurred in gases or liquids, characterized by circulation of fluid around a core. Among all vortical structures, vortex rings (VRs) are perhaps the most familiar one to our daily experience, whose compact and persistent nature has fascinated many researchers for a long time [1]. VRs are vortices whose core is a close loop. They are found to occur in various scales in nature, from the well-known smoke rings of cigarettes to the VRs observed in the wakes of aircraft. Remarkably, the quantized VRs with cores of angstrom size can be formed when charged particles are accelerated through a superfluid helium [2], which were detected in the pioneering work by Rayfield and Reif [3]. Recently, due to the achievement of quantized vortices in the trapped Bose-Einstein condensate (BEC) [4–6], the formation of VRs in ultracold atoms has been highly regarded. Several schemes for producing VRs in atomic BEC have been put forward [7–11]. In particular, Feder *et. al.* [10] have proposed using dynamical instabilities in the condensate to make a dark soliton decay into VRs. Based on this scheme, the VRs in a trapped BEC were first realized experimentally by JILA group [12].

In fluids, VRs can move along their axes with a self-induced velocity. Despite their solitary nature, VRs are susceptible to azimuthally wavy distortions and these so-called bending waves may be amplified under certain circumstances. The bending-wave instability of classical VRs has been extensively studied in the past few decades, and its inviscid instability was first explored by Widnall *et. al.* [13]. This instability is a short-wave instability, characterized by $k\xi \sim O(1)$, where $k$ is the wavenumber of the unstable wave and $\xi$ is the size of vortex core. In this paper, we examined the bending-wave instability of a VR in a trapped condensate. Albeit that the vortex dynamics in a superfluid resembles to that in a normal inviscid fluid in cases where the quantized circulation may be neglected, some essential differences concerning the bending-wave instability remain between these two fluids. For example, the core size of a quantized vortex is extremely small in a large condensate [14], where we may assume $\xi \to 0$. Accordingly, $k \to \infty$ if the condition $k\xi \sim O(1)$ is required. Thus, the Widnall instability does not occur in this limiting case. However, the trapping potential causes vortex stretching, and the bending-wave instability with wavelength comparable to the ring radius may still happen.

To analyze the stability of the VR in a trapped BEC, we use a scheme developed by Svidzinsky and Fetter, which utilizes the quantum analog of Biot-Savart law to determine the local velocity for each element of the vortex [15]. The velocity formula is derived from the time dependent Gross-Pitaevskii equation by using the method of matched asymptotic expansions in the Thomas-Fermi (TF) limit. Considering the axisymmetric potential $V(\mathbf{x}) = m\left(x^2\omega_\perp^2 + y^2\omega_\perp^2 + \omega_z^2 z^2\right)/2$, and defining the aspect ratio $\lambda = \omega_z/\omega_\perp$, the density profile of the condensate in TF limit is given by $\rho(\mathbf{x}) = \rho_0\left(1 - r^2/R_\perp^2 - z^2/R_z^2\right)$ in the cylindrical coordinates $(r, \theta, z)$, where $R_\perp = \left(2\mu/m\omega_\perp^2\right)^{1/2}$ and $R_z = \left(2\mu/m\omega_z^2\right)^{1/2}$ are, respectively, the radial and axial TF radii of trapped BEC; $\mu$ is the chemical potential and $\rho_0 = \mu m/4\pi\hbar^2 a$ is the central particle density. Thus, the velocity of a vortex line element at $\mathbf{x}$ in a non-rotating trap is given by [15]

$$\mathbf{v}(\mathbf{x}) = \Lambda(\xi, \kappa)\left(\kappa\hat{\mathbf{b}} + \frac{\hat{\mathbf{t}} \times \nabla V(\mathbf{x})}{\mu\rho(\mathbf{x})/\rho_0}\right), \qquad (1)$$

where $\Lambda(\xi, \kappa) = (-\hbar/2m)\ln(\xi\sqrt{R_\perp^{-2} + \kappa^2/8})$; $\hat{\mathbf{t}}$ is the unit vector tangent to the vortex line at $\mathbf{x}$, and $\hat{\mathbf{b}}$ is the associated binormal unit vector; $\kappa$ is the curvature of the vortex line at $\mathbf{x}$. Here the vortex is assumed to carry one quantum of circulation.

The first term in the bracket of eq.(1) is from the local induction approximation (LIA), based on the assumption, $\kappa\xi \ll 1$, where the vortex is treated as an infinitely thin filament. It says that the motion is self-induced by the local curvature and heading in the direction of $\hat{\mathbf{b}}$. In fact, LIA alone ensures no vortex stretching since the arc length between any two points on the vortex line remains invariant [1]. Nevertheless, the presence of the external potential indeed gives the stretching force, $\hat{\mathbf{t}} \times \nabla V(\mathbf{x})$. When these two effects are balanced, the VR comes to a stop. Now suppose a circular VR of radius $r$ is formed axisymmetrically in the condensate so that we may initialize $\hat{\mathbf{t}} = \mathbf{e}_\theta$, $\kappa \hat{\mathbf{b}} = (1/r)\mathbf{e}_z$. Now since $\xi \ll r < R_\perp$, the magnitude of the logarithmic term in $\Lambda(\xi,\kappa)$ is predominated by the numerical factor $\ln \xi$, which is large in magnitude as $\xi \to 0$. Therefore $\Lambda(\xi,\kappa)$ varies very slowly with $r$ so that we may treat $\Lambda(\xi,\kappa)$ as a constant. The motion equations then reduce to

$$\frac{1}{\Lambda}\dot{r} = \frac{2\lambda^2 z}{G(r,z)}, \quad \dot{\theta} = 0, \quad \frac{1}{\Lambda}\dot{z} = \frac{-2r}{G(r,z)} + \frac{1}{r}, \quad (2)$$

where $G(r,z) = R_\perp^2 - r^2 - \lambda^2 z^2$. Here we have used the scaled harmonic oscillator units, i.e., the units of length, time and energy are $(\hbar/m\omega_\perp)^{1/2}$, $\omega_\perp^{-1}$, $\hbar\omega_\perp$, respectively. Note that the properties of condensate and trap are specified by $R_\perp$ and $\lambda$ respectively.

The equation $\dot{\theta} = 0$ is trivial, and the motion of VR is solely governed by the other two equations. The numerical solutions of $r$ and $z$ equations above show that the VR moves up and down along $z$ axis cyclically with its radius expanding and shrinking cyclically at the same time. Dividing $\dot{r}$ by $\dot{z}$, we obtain the following equation describing the phase portrait for $r$ and $z$,

$$\frac{dr}{dz} = \frac{2\lambda^2 rz}{R_\perp^2 - 3r^2 - \lambda^2 z^2}, \quad (3)$$

of which the solution is given by $rG(r,z) = C$, where $C$ is a constant. To explore this solution further, we obtain a family of closed contours with different $C$'s for a given $\lambda$, indicating the cyclic motion of the VR. Let $r_{\min}$ and $r_{\max}$ be the minimum and maximum values of $r$ specified by $C$, so we have $C = r_{\min}(R_\perp^2 - r_{\min}^2) = r_{\max}(R_\perp^2 - r_{\max}^2)$. The static solution is given by $r_{eq} = R_\perp/\sqrt{3}$, $z_{eq} = 0$, which is independent of $\lambda$ and agrees with the result in Ref.[8]. Fig.1(a) shows a typical phase portrait of eq.(3) for $\lambda = 1$ and $R_\perp = 10$; Fig.1(b) shows the associated ring's velocity in $z$ direction for various $C$'s. Moreover, let $r(0) = r_{\min}$ for a selected contour and integrate the radial equation in eq.(2), we find that $t$ can be expressed in a closed form of $r$, i.e.,

$$t = \frac{C}{2\Lambda\lambda} \int_{r_{\min}}^{r} \frac{dr}{\sqrt{r^2 R_\perp^2 - r^4 - Cr}} = \frac{C}{2\Lambda\lambda q} F\left(w(r), \frac{p}{q}\right), \quad (4)$$

where $F(\varphi, \eta)$ is the elliptic integral of the first kind [16] with $w(r) = \sin^{-1}[r_{\max}(r - r_{\min})/r(r_{\max} - r_{\min})]^{1/2}$, $p = (r_{\max}^2 - r_{\min}^2)^{1/2}$, $q = [r_{\max}(2r_{\min} + r_{\max})]^{1/2}$. Accordingly, the period of precession along the contour of a selected $C$ is given by $\tau = (C/\Lambda\lambda q) F(\pi/2, p/q)$.

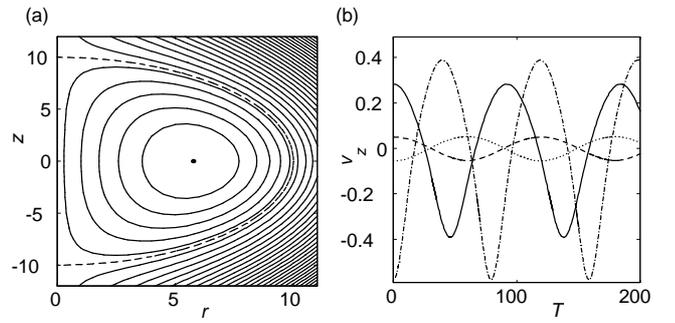

FIG. 1: ((a) Phase portrait for eq.(3) with $\lambda = 1$ and $R_\perp = 10$. The equilibrium position of the ring is located at $r_{eq} = 5.77$ and $z_{eq} = 0$. (b) Propagation velocity along $z$ direction as a function of the scaled time $T = \Lambda t$ for various $r(0)$. The solid, dashed, dotted and dashed-dotted lines represent the cases of $r(0)/r_{eq} = 0.5$, 0.9, 1.1 and 1.5, respectively. The dashed line indicates the TF limit.

To study the bending-wave stability of the base solution given above, let us initialize the coordinates as $r(t) = r_0(t) + \varepsilon r_1(\theta, t)$, $\theta(t) = \theta_0(t) + \varepsilon \theta_1(t)$, and $z(t) = z_0(t) + \varepsilon z_1(\theta, t)$, where $\varepsilon r_1$, $\varepsilon \theta_1$ and $\varepsilon z_1$ denote the small perturbations with $\varepsilon \ll 1$. According to the Seret-Frenet formula, the binormal vector is given by $\kappa \hat{\mathbf{b}} = \kappa \hat{\mathbf{t}} \times \hat{\mathbf{n}} = (\partial \mathbf{x}/\partial s) \times (\partial^2 \mathbf{x}/\partial s^2)$, where $s$ is the arc length. Currently, it is more suitable to write $\hat{\mathbf{t}}$ and $\hat{\mathbf{b}}$ in terms of $\theta$ rather than $s$. Using the chain rule and the relation $ds = |\partial \mathbf{x}/\partial \theta| d\theta$, it follows that $\hat{\mathbf{t}} = (\partial \mathbf{x}/\partial \theta)|\partial \mathbf{x}/\partial \theta|^{-1}$ and $\kappa \hat{\mathbf{b}} = (\partial \mathbf{x}/\partial \theta) \times (\partial^2 \mathbf{x}/\partial \theta^2)|\partial \mathbf{x}/\partial \theta|^{-3}$ with

$$\frac{\partial \mathbf{x}}{\partial \theta} = \varepsilon r_1' \mathbf{e}_r + (r_0 + \varepsilon r_1)\mathbf{e}_\theta + \varepsilon z_1' \mathbf{e}_z + O(\varepsilon^2), \quad (5)$$

$$\frac{\partial^2 \mathbf{x}}{\partial \theta^2} = \left(-r_0 + \varepsilon r_1'' + \varepsilon r_1\right)\mathbf{e}_r + 2\varepsilon r_1' \mathbf{e}_\theta + \varepsilon z_1'' \mathbf{e}_z \quad (6)$$
$$+ O(\varepsilon^2),$$

where the "prime" and "double prime" notations indicate the first and second derivatives with respect to $\theta$. Hence, to the leading order we have

$$\hat{\mathbf{t}} = \frac{\varepsilon r_1'}{r_0}\mathbf{e}_r + \mathbf{e}_\theta + \frac{\varepsilon z_1'}{r_0}\mathbf{e}_z + O(\varepsilon^2), \quad (7)$$

$$\kappa \hat{\mathbf{b}} = \frac{\varepsilon z_1''}{r_0^2}\mathbf{e}_r - \frac{\varepsilon z_1'}{r_0^2}\mathbf{e}_\theta + \frac{r_0 - \varepsilon(r_1 + r_1'')}{r_0^2}\mathbf{e}_z \quad (8)$$
$$+ O(\varepsilon^2).$$

Substituting eqs.(7) and (8) into eq.(1) and expanding all terms in power series of $\varepsilon$, it is easy to show that $(r_0, \theta_0, z_0)$ satisfy the zeroth order equations, eq.(2). To the next order, we get the following linearized equations

$$\frac{\dot{r}_1}{\Lambda} = \frac{z_1''}{r_0^2} + \frac{2\lambda^2 z_1}{G(r_0, z_0)} + \frac{4\lambda^2 z_0(r_0 r_1 + \lambda^2 z_0 z_1)}{G^2(r_0, z_0)}, \quad (9)$$

$$\frac{r_1 \dot{\theta}_0 + r_0 \dot{\theta}_1}{\Lambda} = -\frac{z_1'}{r_0^2} + 2\frac{z_1' - \lambda^2 z_0 r_1'/r_0}{G(r_0, z_0)}, \quad (10)$$

$$\frac{\dot{z}_1}{\Lambda} = -\frac{r_1 + r_1''}{r_0^2} - \frac{2r_1}{G(r_0, z_0)} - \frac{4r_0(r_0 r_1 + \lambda^2 z_0 z_1)}{G^2(r_0, z_0)}. \quad (11)$$

In view of the linear instability for bending waves, we let $r_1(\theta, t) = R_1(t) \exp(in\theta)$, $z_1(\theta, t) = Z_1(t) \exp(in\theta)$ where $n$ is an integer representing the wavenumber. Note that this assumption is valid only when the wavelengths of the bending waves are comparable to the ring radius. As a result, we have

$$\frac{\dot{R}_1}{\Lambda} = \frac{-n^2 Z_1}{r_0^2} + \frac{2\lambda^2 Z_1}{G(r_0, z_0)} + \frac{4\lambda^2 z_0 (r_0 R_1 + \lambda^2 z_0 Z_1)}{G^2(r_0, z_0)}, \quad (12)$$

$$\frac{\dot{Z}_1}{\Lambda} = \frac{(n^2 - 1) R_1}{r_0^2} - \frac{2R_1}{G(r_0, z_0)} - \frac{4r_0 (r_0 R_1 + \lambda^2 z_0 Z_1)}{G^2(r_0, z_0)}. \quad (13)$$

Here we have ignored the equation of $\theta_1$ which is redundant in solving $R_1(t)$ and $Z_1(t)$. In general, $r_0(t)$, $z_0(t)$, $R_1(t)$, and $Z_1(t)$ can only be solved numerically. The linear stability will depend on $\lambda$, $n$ and $r_0(0)$ by letting $z_0(0) = 0$ without loss of generality, and the stability boundaries are shown in Fig.2. If we focus our attention on the linear stability of a static ring, i.e., with $r_0(t) = r_{\rm eq}$, $z_0(t) = z_{\rm eq}$, and further assume $R_1(t) = \alpha \exp(-i\omega t)$, $Z_1(t) = \beta \exp(-i\omega t)$, the condition for nontrivial $\alpha$ and $\beta$ after substituting the expression above into eqs.(12) and (13) is given by

$$\det \begin{bmatrix} -i\omega R_\perp^2 & 3\Lambda(n^2 - \lambda^2) \\ 3\Lambda(n^2 - 3) & i\omega R_\perp^2 \end{bmatrix} = 0 \quad (14)$$

Accordingly, we can obtain the elementary excitation spectrum from eq.(14)

$$\omega_{|n|}(\lambda) = \pm \frac{3\Lambda}{R_\perp^2} \sqrt{(n^2 - \lambda^2)(n^2 - 3)} \quad (15)$$

Obviously, $\omega_{|n|}$ are either purely real or purely imaginary. Real $\omega_{|n|}$ correspond to wavenumbers that oscillate in time with fixed amplitude such that the base solution is stable. If $\omega_{|n|}$ are otherwise imaginary, they occur in complex-conjugate pairs, such that one of them implies the amplitude of bending wave grows exponentially in time, and the base solution is then unstable. Accordingly, we find that when (1) $0 \leq \lambda < 1$, $\omega_{|n|}$ are real for all $|n| \neq 1$; (2) $1 \leq \lambda \leq 2$, $\omega_{|n|}$ are real for all $n$; (3) $\lambda > 2$, $\omega_{|n|}$ are real for $|n| \leq 1$ or $|n| \geq \lambda$. Hence, we conclude that the ring is absolutely stable when $1 \leq \lambda \leq 2$. For other $\lambda$, instability always happens.

The normal mode of the stable wave is given by

$$\begin{bmatrix} r_1^{(n)}(\theta, t) \\ z_1^{(n)}(\theta, t) \end{bmatrix} \propto \begin{bmatrix} 1 \\ \gamma_n \end{bmatrix} e^{i(n\theta - \omega_{|n|} t)} \quad (16)$$

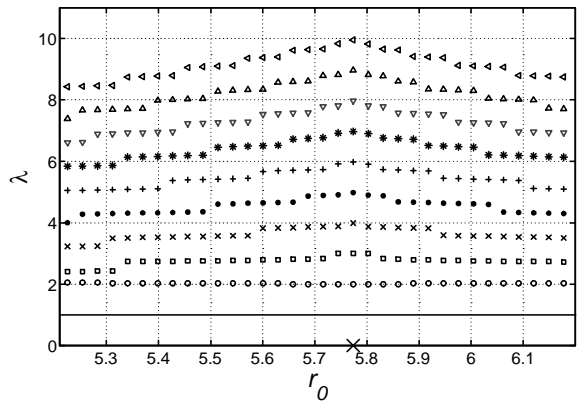

FIG. 2: From bottom to top are stability boundary curves for $n = 1$ to $10$ obtained by solving eqs.(2), (12) and (13) numerically with $R_\perp = 10$. The solid line represents the stability boundary for $n = 1$; the circle for $n = 2$; the square for $n = 3$, and so forth. The region above each boundary curve is unstable for the corresponding mode and stable below for $n = 2$ to $10$. For $n = 1$, with the boundary curve being simply $\lambda = 1$, it is unstable above this boundary and stable below it (and including it). The cross at the buttom of the figure for denotes the equilibrium radius of the ring.

where $\gamma_n = i\omega_{|n|} R_\perp^2 / 3\Lambda (n^2 - \lambda^2)$ is the ratio between $\alpha$ and $\beta$ for the $n$-th mode. The $\pi/2$ phase difference between $\alpha$ and $\beta$ implies a helical wave along the unperturbed circumference of the VR. Since $\omega_{|n|}$ comes in pair as shown in eq.(15), this means that there are two identical helical waves travelling in opposite directions, and the result is a standing wave rotating around unperturbed core axis, which is also called a bending wave. Fig.3 shows the projection of the VR on the plane of $z = 0$ under various stable bending waves. For $n = 0$, the rotation is precisely the precession of a circular ring with $r_0(0)$ very close to $r_{\rm eq}$, and the frequency obtained from eq.(4) by setting $r_{\min} \lesssim r_{eq} \lesssim r_{\max}$ is exactly $\omega_0$. The modes with $|n| = 1$ are very special. Since the ring is slightly shifted from its original equilibrium position, it tilts due to the unbalanced force and then starts wobbling around the $xy$ plane. For $|n| \geq 2$, the bending waves look like a petal.

The normal mode for the unstable wave is obtained by replacing the imaginary frequency into eq.(16). Since $\omega_{|n|}$ is a pure imaginary number, the bending wave, unlike the stable one, is not propagating at all, and solely grows exponentially in time. Moreover, since $\gamma_n$ becomes real for unstable waves, we conclude that $r_1$ and $z_1$ must grow either in phase or out of phase when instability occurs. It turns out that the unstable bending wave grows up in time along the angle of $\tan^{-1} \gamma_n$ relative to the horizontal. The growth of disturbance is shown in Fig.4, where the $n = 8$ mode is initially excited and continues to grow as predicted by eq.(15) under $\lambda = \sqrt{65}$. In most

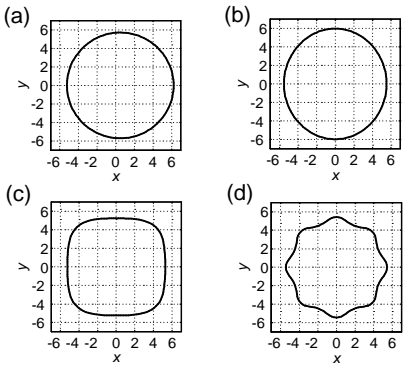

FIG. 3: Projection of a static vortex ring on $z = 0$ plane under various stable normal modes at $\Lambda t = 40$, where $\lambda = 1$ and $R_\perp = 10$. The amplitude for the normal modes is $5 \times 10^{-2} R_\perp$. (a) $n = 1$. (b) $n = 2$. (c) $n = 4$. (d) $n = 8$.

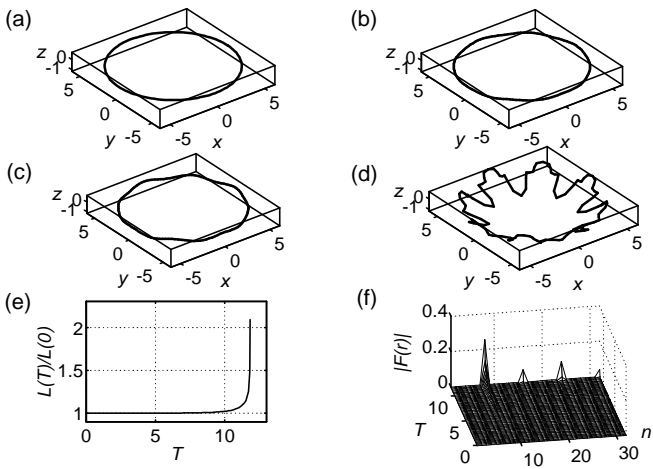

FIG. 4: Growth of unstable waves on a static vortex ring at various scaled time $T = \Lambda t$, where $\lambda = \sqrt{65}$, $R_\perp = 10$. The ring is initially perturbed by the $n = 8$ mode with an amplitude of $5 \times 10^{-4} R_\perp$. (a) $T = 0$. (b) $T = 4.90$. (c) $T = 7.90$. (d) $T = 11.82$. (e) The circumference $L$ of the vortex ring as a function of time. The dynamics is catastrophically disrupt when $L$ diverges, signifying the occurrence of turbulence. (f) Development of the Fourier components of the ring radius. The $n = 8$ mode is most amplified. In the long time limit, higher harmonics with $n = 16, 24$, etc. are amplified due to nonlinear interaction.

situations, however, the perturbation is generally random and only the unstable mode with the largest growth rate will be selected to grow and dominate the instability. This most unstable mode can be found by solving $d(i\omega_n)/dn = 0$. If $n^*$ is the wavenumber of the most unstable mode, then $n^*$ will be closest to $\left(\lambda^2/2 + 3/2\right)^{1/2}$ for $\lambda > 2$. For $0 \leq \lambda < 1$, the only unstable modes are $|n| = 1$ with a growth rate $3\Lambda R_\perp^{-2} \left(2 - 2\lambda^2\right)^{1/2}$ that reaches its maximum at $\lambda = 0$.

In summary, we have investigated the dynamics of a single VR and its stability in an axisymmetric BEC in the TF limit, using a velocity formula derived by Svidzinsky and Fetter [15]. For a unperturbed VR initially placed and oriented axisymmetrically, the motion equations are solved and the associated phase portrait shows that the VR moves up and down accompanied by expanding and shrinking in radius at the same time. The bending-wave instability follows to be studied and the stability boundaries are computed for various combinations of $\lambda$, $n$ and $r_0(0)$. The excitation spectrum featuring the linear stability of a static VR is derived analytically. From this spectrum, an absolute stable region $1 \leq \lambda \leq 2$ is identified, which agrees well with the direct numerical simulation of eq.(1).

This work is supported in part by National Science Council, Taiwan, under contract No. 94-2115-M-002-019.